\documentclass[twocolumn,prl,superscriptaddress,noshowpacs,epsf]{revtex4}

\usepackage{graphicx}

\begin{document}
\title{Low-decoherence flux qubit}
\date{\today}
\author{J. Q. You}
%\altaffiliation[Email address:~jqyou@fudan.edu.cn]{}
\affiliation{Department of Physics and Surface Physics Laboratory (National Key
Laboratory), Fudan University, Shanghai 200433, China}
\affiliation{Frontier Research System, The Institute of Physical
and Chemical Research (RIKEN), Wako-shi 351-0198, Japan}
\author{Xuedong Hu}
%\altaffiliation[Email address:~xhu@buffalo.edu]{}
\affiliation{Department of Physics, University at Buffalo, SUNY,
%The State University of New York,
Buffalo, NY 14260-1500, USA}
\author{S. Ashhab}
%\altaffiliation[Email address:~nori@umich.edu]{}
\affiliation{Frontier Research System, The Institute of Physical
and Chemical Research (RIKEN), Wako-shi 351-0198, Japan}
\author{Franco Nori}
%\altaffiliation[Email address:~nori@umich.edu]{}
\affiliation{Frontier Research System, The Institute of Physical
and Chemical Research (RIKEN), Wako-shi 351-0198, Japan}
\affiliation{Center for Theoretical Physics, Physics Department,
Center for the Study of Complex Systems,
University of
Michigan, Ann Arbor, MI 48109-1040, USA}
%\altaffiliation[Permanent address.]{}

\begin{abstract}
A flux qubit can have a relatively long decoherence time at the degeneracy point,
but away from this point the decoherence time is greatly reduced by dephasing.
This limits the practical applications of flux qubits.
Here we propose a new qubit design modified from the
commonly used flux qubit by introducing an additional capacitor shunted in parallel
to the smaller Josephson junction (JJ) in the loop.
Our results show that the effects of noise can be considerably
suppressed, particularly away from the degeneracy point,
by both reducing the coupling energy of the JJ and increasing the shunt capacitance.
This shunt capacitance provides a novel way to improve the qubit.
\end{abstract}
\pacs{03.67.Pp, 03.65.Yz, 85.25.-j, 03.67.Lx}
\maketitle

Superconducting quantum circuits based on Josephson junctions (JJs) are promising
candidates of qubits for scalable quantum computing (see, e.g., \cite{YN}).
Like other types of superconducting qubits, flux qubits have been shown to
have quantum coherent properties
(see, e.g., \cite{MIT,CHI,YNF,Gra,NTT,Bert,Liu}).
A recent experiment~\cite{Bert} showed that this qubit has a long decoherence
time $T_2$ ($\sim 120$~ns) at the degeneracy point; this $T_2$ can become
as long as $\sim 4$~$\mu$s by means of spin-echo techniques.
However, even {\it slightly away} from the degeneracy point, the decoherence
time is drastically reduced.  This sensitivity to flux bias
considerably limits the applications both for
flux qubits for quantum computing, and also when performing quantum-optics
and atomic-physics experiments on microelectronic chips with the qubit as an
artificial atom.

Typically, JJ circuits have two energy scales: the charging
energy $E_c$ of the JJ, and the Josephson coupling energy $E_J$
of the junction.  Ordinarily, a flux qubit works in the phase regime with
$E_J/E_c\gg 1$, where its decoherence is dominated by flux fluctuations.
For the widely used three-junction flux qubit design~\cite{MIT,CHI,YNF,Gra,NTT},
in addition to two identical JJs with coupling energy $E_J$ and
charging energy $E_c$, a third JJ, which has an area smaller by a factor
$\alpha\sim 0.7$, is employed to properly adjust the qubit spectrum.
Charge fluctuations can affect the decoherence of this flux qubit
via the smaller junction.

Here we search for an improved design for flux qubits.
We show that reducing the ratio $E_J/E_c$ suppresses the effects of flux
noise, although charge noise becomes increasingly important.  Reducing $\alpha$
further suppresses the effects of flux noise and considerably improves the
decoherence properties away from the degeneracy point.  As the effect of
flux noise has been largely suppressed, charge noise would now be the dominant
source of decoherence. It mainly comes from the charge fluctuations on the
two islands separated by the smaller JJ and affects the qubit mainly through
relaxation.
We thus propose an improved flux qubit by introducing a large capacitor
that shunts in parallel to the smaller JJ.  This shunt capacitance suppresses
the effects of the dominant charge noise in the two islands separated by the
smaller JJ by reducing the charging energy.
Our results reveal that using a larger shunt capacitor allows reducing both
$E_J/E_c$ and $\alpha$ to considerably suppress the effects of both flux and
charge noises, particularly away from the degeneracy point.
In essence, our method reduces the couplings of the flux qubit to the {\it two}
types of noise.  It provides a promising approach for
lowering the decoherence of
%superconducting
JJ qubits.

\begin{figure}
\includegraphics[width=2.2in,  %%% TWO-COLUMN 4-PAGE VERSION FOR ARXIV
bbllx=12,bblly=281,bburx=555,bbury=623]{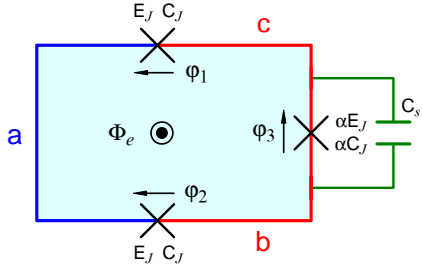}
\caption{(Color
online) (a)~Schematic diagram of a low-decoherence flux qubit.
The loop is pierced by an externally applied magnetic
flux $\Phi_e$ and interrupted by three Josephson junctions (JJs) that
divide the loop into three islands denoted by $a$ (blue), $b$ (red)
and $c$ (red). The two identical JJs have
coupling energy $E_J$ and capacitance $C_J$, while the third (smaller) JJ
has coupling energy $\alpha E_J$ and capacitance $\alpha C_J$,
where $0.5<\alpha<1$. Here a capacitance $C_s$ is shunted in parallel to
the smaller JJ, so as to reduce the charging energy related to islands $b$ and $c$.
The shunt capacitor can be fabricated by depositing a metallic plate
below or above the smaller junction, with a dielectric layer between the metallic
plate and $b$ as well as between the plate and $c$.}
%Usually, a flux qubit
%has a loop size $\sim 1$~$\mu$m. As shown in \cite{YNF}, when the loop
%size increases to $\sim 100$~$\mu$m, the flux qubit is not significantly
%affected by the self inductance of the loop. Thus, one can employ a flux
%qubit with a relatively large loop, so as to obtain a large shunt $C_s$ by producing
%large-area oxidized layers.}
\label{fig1}
\end{figure}

We consider, as a typical example, a modified version of
the usual three-junction flux qubit~\cite{MIT,CHI,YNF,Gra,NTT};
it consists of a superconducting loop containing three JJs and pierced by a
magnetic flux.  The new ingredient, which drastically improves the qubit, is an
additional capacitance $C_s$ shunted in parallel to the smaller JJ (Fig.~\ref{fig1}).
Note that our approach also applies to flux qubits with any number of junctions,
e.g., the four-junction design, by shunting a capacitor to the smaller junction.
As shown in Fig.~\ref{fig1},
the three JJs divide the superconducting loop into three islands,
denoted by $a$, $b$, and $c$.  When the environment-induced charges
on the three islands are taken into account, the Hamiltonian of the system is
%
%\begin{equation}
$H=E_p(n_p-\delta\! N_a)^2+E_m\left[n_m-(\delta\! N_b-\delta\! N_c)\right]^2
+U(\varphi_p,\varphi_m)$,
%\end{equation}
%
where $n_k=-i\partial/\partial\varphi_k$ ($k=p,m$),
%
%$\varphi_p=(\varphi_1+\varphi_2)/2$, and $\varphi_m=(\varphi_1-\varphi_2)/2$;
%
$E_p=2E_c$, and $E_m=E_p/(1+2\beta)$,
with $E_c=e^2/2C_J$, and $\beta=\alpha+C_s/C_J$.
Here $E_p$ is the charging energy of the island $a$, $E_m$ is the
effective charging energy related to islands $b$ and $c$, and  $\delta\! N_i$
($i=a$, $b$ and $c$) are the environment-induced charges (in units of $2e$)
on the islands. The potential energy
%$U(\varphi_p,\varphi_m)$
is
%
%\begin{eqnarray}
%&&U(\varphi_p,\varphi_m)=2E_J(1-\cos\varphi_p\cos\varphi_m) \nonumber\\
%&&~~~~~~~~~~~~~~~~~+\alpha E_J[1-\cos(2\pi f+2\varphi_m)],
%
$U(\varphi_p,\varphi_m)=2E_J(1-\cos\varphi_p\cos\varphi_m)
+\alpha E_J[1-\cos(2\pi f+2\varphi_m)]$,
%\end{eqnarray}
%
%\begin{equation}
%U=2E_J(1-\cos\varphi_p\cos\varphi_m)
%+\alpha E_J[1-\cos(2\pi f+2\varphi_m)],
%\end{equation}
%
where $\varphi_p=(\varphi_1+\varphi_2)/2$, and $\varphi_m=(\varphi_1-\varphi_2)/2$.
The reduced flux $f$ is given by $f=f_e+\delta\! f$, with $f_e=\Phi_e/\Phi_0$
and $\delta\! f=\delta\Phi/\Phi_0$, where $\Phi_e$ is the externally applied
magnetic flux in the loop, $\delta\Phi$
the flux fluctuations, and $\Phi_0=h/2e$.
%The number operator $n_k$ of the Cooper pairs and the phase $\varphi_j$
%obey $[\varphi_j,n_k]=i\delta_{jk}$, where $j,k=p,m$.

We study the environment-induced effects on the qubit by expanding the
Hamiltonian $H$ into a series and truncating it to keep the interaction terms
to lowest order.  Including the environmental Hamiltonian
$H_{\rm env}$, we can write the total Hamiltonian $H_t$ as
$H_t=H_q+H_{\rm env}+H_{\rm int}$, where $H_q=H(\delta\! N_i=0,\delta\! f=0)$
is the Hamiltonian of the qubit and the interaction Hamiltonian takes the form
%
%\begin{equation}
$H_{\rm int} = -2E_p\, n_p\,\delta\! N_a-2E_m\, n_m(\delta\! N_b - \delta\! N_c)
-I\,\Phi_0\,\delta\! f$,
%\end{equation}
%
where $I=-\alpha\, I_c\sin(2\pi f_e + 2\varphi_m)$ with $I_c=2\pi E_J/\Phi_0$.
%
%is the circulating supercurrent of the loop in the absence of
%the flux noise $\delta f$.

To study the coherence properties of this flux qubit, we project the total
Hamiltonian onto the subspace spanned by the qubit eigenstates $|0\rangle$
and $|1\rangle$ with eigenenergies $E_0$ and $E_1$,
the two lowest energy levels of the quantum device.
%This projection procedure is consistent with our truncation at the lowest order
%for the interaction Hamiltonian.
Now the flux qubit Hamiltonian is reduced to
%\begin{equation}
$H_q=\frac{1}{2}\varepsilon \sigma_z$,
%\end{equation}
with $\varepsilon=E_1-E_0$ and $\sigma_z=|1\rangle\langle 1|-|0\rangle\langle 0|$;
while the interaction Hamiltonian is reduced to
\begin{equation}
H_{\rm int}= - \sum_i \left[\;\sigma_z X_i - (\sigma_ + Y_i + {\rm H.c.}) \right],
\label{Hint}
\end{equation}
with $\sigma_+=|1\rangle\langle 0|$.
%being the raising operator.
Here the longitudinal couplings $X_i$, $i=f$, $a$, $b$ and $c$, are
%have the form of
$X_i(t)=A_i\,\delta\! \lambda_i(t)$, with
$A_f=\frac{1}{2}\Phi_0\left(\langle 1|I|1\rangle-\langle 0|I|0\rangle\right)$,
$A_a=E_p\left(\langle 1|n_p|1\rangle-\langle 0|n_p|0\rangle\right)$,
and $A_b=-A_c=E_m\left(\langle 1|n_m|1\rangle-\langle 0|n_m|0\rangle\right)$.
The transverse couplings $Y_i$ are $Y_i(t)=B_i\,\delta\! \lambda_i(t)$, where
$B_f=\Phi_0 \langle 1|I|0\rangle$, $B_{a}=2E_p \langle 1|n_p|0\rangle$,
and $B_b=-B_c=2E_m \langle 1|n_m|0\rangle$.
The fluctuations are $\delta\! \lambda_f\equiv\delta\! f$ for the flux noise
and $\delta\! \lambda_i\equiv\delta\! N_i$ ($i=a$, $b$, and $c$)
for the charge noises related to the three islands.
The longitudinal coupling term $\sigma_z X_i$ leads to pure dephasing between
the qubit states,
while the transverse coupling term, $\sigma_+Y_i +\rm H.c.$, leads to relaxation.
%%Because $X_i(t)=A_i\,\delta \lambda_i(t)$, and $Y_i(t)=B_i\,\delta\lambda_i(t)$,
%
One way to suppress decoherence from {\it both} pure dephasing and relaxation
is to reduce the longitudinal and transverse couplings by
decreasing $|A_i|$ and $|B_i|$.  This general method of decoherence suppression
applies {\it irrespective} of the particular behavior of $\lambda_i(t)$, i.e.,
whether it is Gaussian or non-Gaussian noise.
%
%Below we calculate the quantities characterizing the longitudinal and transverse
%coupling strengths and compare the relative contributions by flux and charge noises.

\begin{figure}
\includegraphics[width=3.0in, %% Two-columns width figure 4-PAGE VERSION FOR ARXIV
bbllx=29,bblly=92,bburx=574,bbury=740]{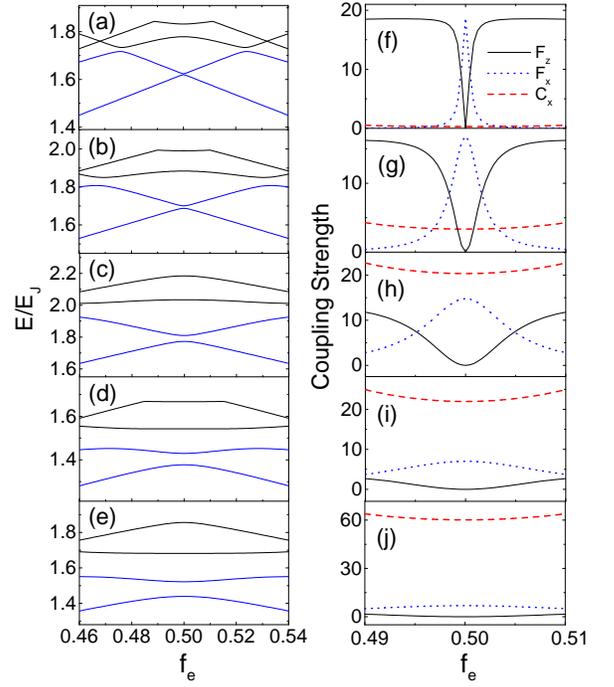}
\caption{(Color online) (a)-(e):~Energy levels of the flux qubit
versus the reduced magnetic flux $f_e=\Phi_e/\Phi_0$
for $\alpha=\beta=0.8$ and $E_J/E_c=$ (a)~$60$, (b)~$35$ and (c)~$20$
as well as $\alpha=\beta=0.6$ and $E_J/E_c=$ (d)~$60$ and (e)~$35$.
Here only the lowest four energy levels are shown.
(f)-(j):~Coupling strength (in units of $E_J^2$)
versus the reduced flux $f_e$, which correspond to (a)-(e), respectively.
%Note that $C_{x\Gamma}=C_{x\Lambda}\equiv C_x$.
In this figure, and the following one, other coupling strength parameters
for charge noise are not shown because they are orders of magnitude smaller.
}
\label{fig2}
\end{figure}

We first study the conventional three-junction flux qubit,
without the shunt capacitance $C_s$.
We take $\alpha=\beta=0.8$, as in the experiment in Ref.~\cite{CHI}.
Figures~\ref{fig2}(a)-\ref{fig2}(c) show the flux dependence of the energy levels.
The lowest two levels around the degeneracy point $f_e \equiv \Phi_e/\Phi_0=0.5$
are employed as the qubit states.  To characterize the effects of the flux noise
on this qubit, we introduce longitudinal and tranverse coupling strengths defined by
$F_z=|A_f|^2$ and $F_x=|B_f|^2$, respectively.
To compare the contributions of charge noise
with flux noise, we define the longitudinal and transverse coupling
strengths $C_{zi}=\kappa_i\, |A_i|^2$ and $C_{xi}=\kappa_i\, |B_i|^2$
for the charge noise, where $i=a$, $b$ and $c$.  The coefficient
$\kappa_i$ characterizes the relative contribution of each charge noise and
is defined as $\kappa_i=S_i(\omega)/S_f(\omega)$,
i.e., the ratio between the power spectra of each charge noise and the flux noise.
This definition is reasonable as the qubit relaxation rate is proportional to
both $|B_i|^2$ and the power spectrum of the noise, as shown below
[see Eq.~(\ref{tran})].
%
%Because $1/f$ noise may be the main source of decoherence,
%
We estimate $\kappa_i$ by considering the $1/f$ noise with power spectrum
$K_i/|\omega|$. Here $K_i$ is determined from experiments.  Typically,
$K_{\rm charge}=(0.3\times 10^{-3})^2$ for the charge noise~\cite{NEC1} and
$K_{\rm flux}=3\times 10^{-12}$ for the flux noise~\cite{Bert}.
For simplicity, the same $K_{\rm charge}$ is used for the three charge noises
related to the islands $a$, $b$ and $c$,
so that $\kappa_i=K_{\rm charge}/K_{\rm flux}\equiv \kappa$.

In Figs.~\ref{fig2}(f)-\ref{fig2}(h), we show the flux dependence of the coupling
strengths $F_z$, $F_x$, and $C_x$.  At $f_e=0.5$, $F_z$ falls to zero, while $F_x$
rises to its peak. This implies that, at the degeneracy point $f_e=0.5$,
the first-order pure dephasing due to flux noise disappears, and the flux qubit
decoherence is dominated by relaxation.  However, for a large $E_J/E_c$
[see, e.g., Fig.~\ref{fig2}(f)], pure dephasing
dominates when $f_e$ is slightly away from the degeneracy point.
For decreasing $E_J/E_c$, the valley of $F_z$ around $f_e=0.5$ becomes broader,
while the peak of $F_x$ becomes less sharp and its height is gradually reduced.
In Figs.~\ref{fig2}(f)-\ref{fig2}(h) we also show the dominant
$C_{xb}=C_{xc}\equiv C_x$ curves due to charge fluctuations on the smaller
islands $b$ and $c$.
The quantity $C_x$ characterizes qubit relaxation induced by charge noise and should
be compared to the $F_x$ curves in the figures.  Notice that when $E_J/E_c$ decreases,
charge noise plays an increasingly important role and eventually is more
important than flux noise in terms of relaxation [compare the dashed and dotted
curves in Fig.~\ref{fig2}(h)].
We note here that as shown in \cite{MIT}, the energy levels of the flux qubit are
very flat versus the offset charges.  This insensitivity of the energy splittings to
the charge fluctuations implies a very weak pure dephasing caused by charge noise.
Indeed, we numerically calculated the longitudinal coupling strengths $C_{za}$
and $C_{zb}=C_{zc}\equiv C_z$ for our system,
and found that these quantities are orders of magnitude smaller than $C_x$.
These results further reveal that the charge-noise-induced dephasing is weak.

Now we reduce the size of the smaller JJ in the conventional 3-JJ flux qubit
to $\alpha=\beta=0.6$. We also show the flux dependence of the energy levels
[Figs.~\ref{fig2}(d) and \ref{fig2}(e)] and the coupling strengths
[Figs.~\ref{fig2}(i) and \ref{fig2}(j)].  Here the coupling
strengths $F_z$ and $F_x$ are reduced and become flatter around $f_e=0.5$.
This reduction of the flux-noise effects is due to the decrease of the
circulating current $I$ with $\alpha$.
%which reduces the coupling of the qubit with the flux noise.
%
However, the coupling strength $C_x$ becomes much larger than $F_z$ and $F_x$
because the effects of charge noise are strengthened due to the increase of
both charging energy $E_m$ and transition matrix element $|\langle 1|n_m|0\rangle|$.
%as $\beta$ decreases.
The results here reveal that the {\it decoherence} of the flux qubit is
{\it sensitive} to the values of $\alpha$ and $\beta$ of the small JJ
(here $\alpha=\beta$). Moreover, the coupling strength $C_x$
increases rapidly with decreasing $E_J/E_c$ [see the dashed curves in
Figs.~\ref{fig2}(i) and \ref{fig2}(j)].  In short, decreasing $\alpha$ reduces the
coupling strength between the flux qubit and the flux environment, and makes
the coupling strength less sensitive to the flux bias so that pure dephasing
becomes less important near the degeneracy point.  However, decreasing $\alpha$
also leads to a dramatic increase in the coupling strength between
the flux qubit and its charge environment, to the degree that it may become
the dominant decoherence channel.

\begin{figure}
 \includegraphics[width=2.9in, %% Two-columns width figure 4-PAGE VERSION FOR ARXIV
bbllx=20,bblly=197,bburx=560,bbury=556]{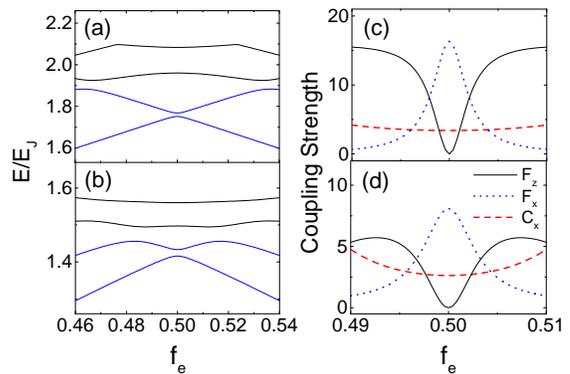}
\caption{(Color online) (a) and (b):~Energy levels of the flux qubit
versus the reduced magnetic flux $f_e$.
Here, (a)~$\alpha=0.8$, $\beta=1.5$ and $E_J/E_c=20$;
(b)~$\alpha=0.6$, $\beta=4$ and $E_J/E_c=35$.
(c) and (d):~Coupling strength (in units of $E_J^2$)
versus the reduced flux $f_e$, which correspond to (a) and (b), respectively.
}
\label{fig3}
\end{figure}

%In order
To achieve an improved flux qubit in which the effects of both
charge and flux noises are reduced significantly,
we shunt a large capacitance in parallel to the smaller JJ (see Fig.~\ref{fig1})
so as to decrease the charging energy $E_m$ while keeping the ratio $\alpha$ small.
In Fig.~\ref{fig3} we present two examples in which the effects of
the charge noise are reduced.  In both cases, a small and flat $C_x$ is achieved.
Also, we show that the coupling strengths $F_z$ and $F_x$ are smaller
and flatter in Fig.~\ref{fig3}(d) than in Fig.~\ref{fig3}(c).
These results indicate that for a suitably chosen $E_J/E_c$ ratio, by optimally
decreasing $\alpha$ and increasing $\beta$ one can reduce the coupling of the
qubit to {\it both} flux and charge noises, so that pure dephasing can be
considerably reduced in a {\it wide region} around the degeneracy point $f_e=0.5$
and the relaxation is significantly suppressed.
This corresponds to an improved flux qubit with low decoherence.

Note that the parameter $\alpha$ has a lower bound of 0.5 for the
flux qubit; when $\alpha<0.5$, the double-well potential reverts
back to a single-well potential and the circuit behaves like a
phase qubit. Also, the shunt capacitance should have an upper
bound when other factors are taken into account.  For instance, a
very large shunt capacitance needs a thicker dielectric insulator
for fabricating the external capacitor. In this case, the
decoherence originating from phonon radiation~\cite{Ioffe} and
defects in the thicker insulator~\cite{MAR} may play more
important roles. Furthermore, when the shunt capacitance
increases, the energy gap $\Delta$ of the qubit at the degeneracy
point narrows down, raising the single-qubit operation time
$h/\Delta$.  This decrease in $\Delta$ is another factor one needs
to consider in determining the upper bound of the shunt
capacitance. One should keep in mind here that reducing $\alpha$
increases $\Delta$. As a result, with properly chosen values of
$\alpha$ and $\beta$, the gap $\Delta$ is not necessarily
decreased in the optimized design, as we shall show with an
example below.

%Moreover, $\alpha$ remains unchanged when increasing
%$\beta$, so the coupling between the driving microwave field and the flux qubit
%is not afftected, for a given field intensity.

\begin{figure}
 \includegraphics[width=2.9in, %% Two-columns width figure 4-PAGE VERSION FOR ARXIV
bbllx=51,bblly=206,bburx=564,bbury=488]{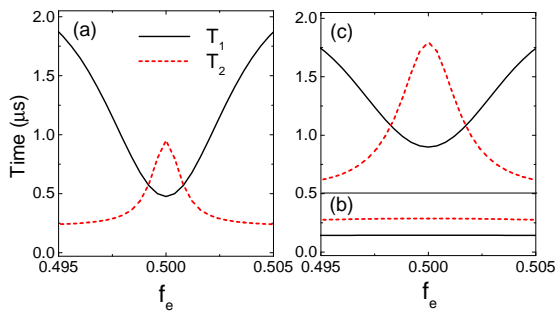}
\caption{(Color online) Relaxation and decoherence times ($T_1$ and $T_2$)
for three flux qubits with the same noise sources. The parameters are $E_J/E_c=35$,
$\alpha=\beta=$~(a)~$0.8$, (b)~$0.6$, and (c)~$\alpha=0.6$, $\beta=4$.
Here, as a simple estimation, we ignore the frequency dependence of the power spectra
around $f_e=0.5$ and determine their high- and low-frequency components
via Eqs.~(\ref{tran}) and (\ref{deph}) by assuming $T_1=0.5~\mu$s and $T_2=0.8~\mu$s
at $f_e=0.5004$ for the qubit in (a).
}
\label{fig4}
\end{figure}

Finally, we discuss the relation between the decoherence rate and our defined
coupling strengths.
When each transverse coupling term in Eq.~(\ref{Hint}) is treated as a perturbation,
according to the Fermi golden rule, one can obtain the ($|B_i|$ dependent)
relaxation rate for each noise:
\begin{equation}
\Gamma_1^{(i)}=\frac{1}{\hbar^2}\,|B_i|^2\, S_i(\omega_{10}),
\label{tran}
\end{equation}
where $\omega_{10}=(E_1-E_0)/\hbar$, and the power spectrum is defined by
$S_i(\omega)=\int_{-\infty}^{+\infty}dt
\langle \lambda_i(t+\tau)\lambda_i(t)\rangle e^{-i\omega\tau}$.
%The relaxation rate $\Gamma_1^{(i)}$ is proportion to $K_i|B_i|^2$ for the $1/f$
%noise $S_i(\omega)=K_i/|\omega|$.
The sum of all $\Gamma_1^{(i)}$ gives the total relaxation rate
$\Gamma_1=1/T_1$, where $T_1$ is the relaxation time.
%
%$T_1=1/\Gamma_1$ characterizes the
%environment-induced damping of the diagonal density matrix elements of the qubit.

The longitudinal qubit-environment coupling introduces a random phase between the
qubit eigenstates.  At time $\tau$ this random phase is
$\Delta\phi_i=(1/\hbar)\int_0^{\tau}dt X_i(t)$.  For a Gaussian noise,
%
%the random phase $\Delta\phi_i$ is Gaussian distributed and
%
the dephasing factor $\eta_i(\tau)$ is given by \cite{NEC1}:
$e^{-\eta_i(\tau)}\equiv\langle e^{i\Delta\phi_i}\rangle
=\exp[-\frac{1}{2}\langle\left(\Delta\phi_i\right)^2\rangle]$,
%
%=\exp\left[-\frac{1}{2\hbar^2}
%\left\langle\left(\int_0^{\tau} dt X_i(t)\right)^2\right\rangle\right],
%\end{equation}
%
where the brackets $\langle\dots\rangle$ denote the quantum statistical average
over the environment. Because $X_i(t)=A_i\,\delta\! \lambda_i(t)$,
one can write the dephasing factor as
\begin{equation}
\eta_i(\tau)=\frac{1}{\hbar^2}\,|A_i|^2
\int_{\omega_c}^{+\infty}\!\!d\omega \, S_i(\omega)
\frac{\sin^2(\omega\tau/2)}{2\pi(\omega/2)^2},
\label{deph}
\end{equation}
where $\omega_c$ is a low-frequency cutoff determined by the measurement time.
%of the experiment.
%The dephasing factor is proportional to $K_i |A_i|^2$ for the $1/f$ noise.
The pure dephasing rate $\Gamma_{\varphi}=1/T_{\varphi}$ is defined
by $\sum_i\eta_i(T_{\varphi})=1$.
%For the $1/f$ noise $S_i(\omega)=K_i/|\omega|$, when $\tau \ll 1/\omega_c$,
%the dephasing factor is reduced to~\cite{Saclay}:
%$\eta_i(\tau)=(\tau^2/2\pi\hbar^2)K_i|A_i|^2\ln(1/\omega_c\tau)$.
The dephasing factor $\eta_i(\tau)\propto |A_i|^2$,
consistent with reducing pure dephasing by decreasing the longitudinal coupling.
%
%This should apply to the non-Guanssian noise; at least, the leading term of the
%dephasing factor should be so.
Moreover, relaxation can also cause damping of the off-diagonal density
matrix elements. Following the Bloch-Redfield
theory (see, e.g., \cite{Geva}), the total damping of the off-diagonal
elements is characterized by a decoherence rate $\Gamma_2=1/T_2$, with
%\begin{equation}
$\Gamma_2=\frac{1}{2}\Gamma_1+\Gamma_{\varphi}$.
%\end{equation}

In Fig.~\ref{fig4}, we show $T_1$ and $T_2$ for three hypothetical flux qubits 
calculated using the same noise sources. The qubit
in Fig.~\ref{fig4}(a), with $\alpha=\beta=0.8$, is 
a conventional qubit with no shunt capacitance~\cite{CHI}. The
qubits in Figs.~\ref{fig4}(b) and \ref{fig4}(c) have a reduced
value of $\alpha$ (0.6 for both figures), and a shunt capacitor (with $\beta=4$) 
is added for Fig.~\ref{fig4}(c). As compared with the conventional
qubit in Fig.~\ref{fig4}(a), the decoherence time $T_2$ in Fig.~\ref{fig4}(c) 
is larger by a factor of 2 at the degeneracy point,  and the $T_2$ peak 
is broader. 
%For instance, at $f_e=0.5$,
%$T_2$ is improved by about a factor 2 in Fig.~4(c) as compared with Fig.~4(a).
%However, when $f_e$ is slightly off this point,
%$T_2$ is improved by about a factor 3; e.g., at $f_e=0.4988$, it is increased from
%$T_2\approx 0.44$~$\mu$s in Fig.~4(a) to $T_2\approx 1.32$~$\mu$s in Fig.~4(c).
%Moreover, the gap $\Delta$ at $f_e\sim 0.5$ is
%about $0.013E_J$ for the qubit in Fig.~4(a), but it is increased to $0.02E_J$
%for the new qubit in Fig.~4(c).
%Thus, the quality factor of quantum coherence
%$Q_{\varphi}\equiv \pi\nu_{01}T_2=\pi(\Delta/h)T_2$ for the new qubit
%is improved by about a factor 3
%at $f_e=0.5$ and by about a factor 5 when $f_e$ is slightly off this point
%(e.g., at $f_e=0.4988$).
%This reveals that the quality of coherence can be significantly improved in
%the new qubit with shunt capacitance.
Furthermore, the qubit in Fig.~\ref{fig4}(c) has a larger gap than
the qubit in Fig.~\ref{fig4}(a), $\Delta\approx 0.02 E_J$ compared
to $\Delta\approx 0.013 E_J$. Combining the increase in both $T_2$ and
$\Delta$, we find that the quality of quantum
coherence is improved by a factor of 3 at the degeneracy point.
Away from the degeneracy point (e.g., at $f_e=0.4988$), $T_2$ can
be improved by a factor of 3 and the quality of coherence is
improved by a factor 5 (nearly one order of magnitude).
%Thus, the new qubit with shunt capacitance has a significantly improved
%quality of quantum coherence.
Also, at $f_e\sim 0.5$, $T_2$ is reduced by about one order of
magnitude if the shunt capacitor is removed [comparing
Figs.~\ref{fig4}(c) and \ref{fig4}(b)]. These results further show
the important role of the shunt capacitance in achieving a
low-decoherence flux qubit.

In conclusion, we have proposed a new qubit design modified from the
commonly used flux qubit. The qubit decoherence is reduced by shunting
the small JJ with an additional capacitor. We show that by increasing the shunt
capacitance and reducing the coupling energy of the JJ, the effects of
both charge and flux noises are considerably suppressed.
%particularly at those points away from the degeneracy point.
Recently, a shunt capacitor was used to improve the performance of phase
qubits~\cite{Steffen}. In that case the motivation for adding the shunt capacitor
was quite different from ours; they used a smaller junction so as to reduce
the number of two-level systems in the junction but decreased the charging
energy of the junction with a shunt capacitor in order to push the qubit back into
the phase regime. However, in our case the effects of the noise are suppressed
even though we assume that the noise source remains unchanged.

We would like to thank M. Grajcar and A.M. Zagoskin for valuable discussions.
This work was supported in part by
the NSA, LPS and ARO.
%the National Security Agency,
%Laboratory of Physical Sciences,
%and the Army Research Office.
J.Q.Y. was supported by the NSFC
%National Natural Science Foundation of China
grant Nos.~10474013, 10534060 and 10625416. S.A. was supported by the JSPS.


\vspace*{-0.19in}

\begin{references}
%
\vspace*{-0.19in}
%
\bibitem{YN} J.Q. You and F. Nori, Phys. Today {\bf 58}(11), 42 (2005).
%
%\bibitem{Mooij} J.E. Mooij {\it et al.}, Science {\bf 285}, 1036 (1999).
%
\bibitem{MIT} T.P. Orlando {\it et al.}, Phys. Rev. B {\bf 60}, 15398 (1999).
%
\bibitem{CHI} I. Chiorescu {\it et al.}, Science {\bf 299}, 1869 (2003).
%
\bibitem{YNF} J.Q. You, Y. Nakamura, and F. Nori,
Phys. Rev. B {\bf 71}, 024532 (2005).
%
\bibitem{Gra} E. Il'ichev {\it et al.}, Phys. Rev. Lett. {\bf 91}, 097906 (2003);
M. Grajcar {\it et al.}, {\it ibid.} {\bf 96}, 047006 (2006).
%
\bibitem{NTT} S. Saito {\it et al.}, Phys. Rev. Lett. {\bf 93}, 037001 (2004);
F. Yoshihara {\it et al.}, {\it ibid.} {\bf 97}, 167001 (2006);
B.L.T. Plourde {\it et al.}, Phys. Rev. B {\bf 72}, 060506 (2005).
%
\bibitem{Bert} P. Bertet {\it et al.}, Phys. Rev. Lett. {\bf 95}, 257002 (2005);
cond-mat/0412485 contains additional information.
%
\bibitem{Liu} Y.X. Liu {\it et al.}, Phys. Rev. Lett. {\bf 96}, 067003 (2006).
%
\bibitem{NEC1} Y. Nakamura {\it et al.}, Phys. Rev. Lett. {\bf 88}, 047901 (2002).
%
\bibitem{Ioffe} L.B. Ioffe {\it et al.}, Phys. Rev. Lett. {\bf 93}, 057001 (2004).
%
\bibitem{MAR} J.M. Martinis {\it et al.}, Phys. Rev. Lett. {\bf 95}, 210503 (2005).
%
%\bibitem{Saclay} G. Ithier {\it et al.}, Phys. Rev. B {\bf 72}, 134519 (2005).
%
\bibitem{Geva} C. Cohen-Tannoudji {\it et al.},
%J. Dupont-Roc, and G. Grynberg,
{\it Atom-Photon Interaction}
%:~Basic Process and Applications}
(Wiley, New York, 1992), Chap. 4.
%\bibitem{Geva}
%E. Geva {\it et al.},
%R. Kosloff, and J.L. Skinner,
%J. Chem. Phys. {\bf 102}, 8541 (1995).
%
\bibitem{Steffen} M. Steffen {\it et al.}, Phys. Rev. Lett. {\bf 97}, 050502 (2006).
%
%\bibitem{Pala} E. Paladino {\it et al.}, Phys. Rev. Lett. {\bf 88}, 228304 (2002).
%
%\bibitem{GAS} Y.M. Galperin, B.L. Altshuler, D.V. Shantsev, cond-mat/0312490.
%
%\bibitem{Faoro} L. Faoro {\it et al.}, Phys. Rev. Lett. {\bf 95}, 046805 (2005);
%O. Astafiev {\it et al.}, Phys. Rev. Lett. {\bf 93}, 267007 (2004).
%

\end{references}
\end{document}